\documentclass[aps,prl,reprint,superscriptaddress,nofootinbib]{revtex4-2}

\usepackage{amsmath,amssymb,graphicx,bm,multirow}
\usepackage[colorlinks=true,citecolor=blue,linkcolor=blue,urlcolor=blue]{hyperref}
\usepackage{booktabs}
\usepackage{soul}
\usepackage{xcolor}

\begin{document}

\title{Extended High-Mass Axion Search with an Auxetically Tuned Higher-Order-Mode Cavity}

\author{Jinsu Kim}
\thanks{These authors contributed equally to this work.}
\affiliation{Dark Matter Axion Group, Institute for Basic Science, Daejeon, Republic of Korea}
\affiliation{Center for Axion and Precision Physics Research, Institute for Basic Science, Daejeon, Republic of Korea}
\author{Sungjae Bae}
\thanks{These authors contributed equally to this work.}
\affiliation{Dark Matter Axion Group, Institute for Basic Science, Daejeon, Republic of Korea}
\affiliation{Center for Axion and Precision Physics Research, Institute for Basic Science, Daejeon, Republic of Korea}
\affiliation{RIKEN Center for Quantum Computing, Wako, Saitama, Japan}
\author{Junu Jeong}
\affiliation{Center for Axion and Precision Physics Research, Institute for Basic Science, Daejeon, Republic of Korea}
\affiliation{Oskar Klein Centre, Department of Physics, Stockholm University, Stockholm, Sweden}
\author{Younggeun Kim}
\affiliation{Center for Axion and Precision Physics Research, Institute for Basic Science, Daejeon, Republic of Korea}
\affiliation{Johannes Gutenberg-Universit\"at Mainz, Mainz, Germany}
\affiliation{Helmholtz Institute Mainz, Mainz, Germany}
\author{Jihn E. Kim}
\affiliation{Department of Physics, Seoul National University, Seoul, Republic of Korea}
\author{Arjan F. van Loo}
\affiliation{RIKEN Center for Quantum Computing, Wako, Saitama, Japan}
\affiliation{Department of Applied Physics, Graduate School of Engineering, The University of Tokyo, Tokyo, Japan}
\author{Yasunobu Nakamura}
\affiliation{RIKEN Center for Quantum Computing, Wako, Saitama, Japan}
\affiliation{Department of Applied Physics, Graduate School of Engineering, The University of Tokyo, Tokyo, Japan}
\author{Seonjeong Oh}
\affiliation{Dark Matter Axion Group, Institute for Basic Science, Daejeon, Republic of Korea}
\affiliation{Center for Axion and Precision Physics Research, Institute for Basic Science, Daejeon, Republic of Korea}
\author{Taehyeon Seong}
\affiliation{Dark Matter Axion Group, Institute for Basic Science, Daejeon, Republic of Korea}
\affiliation{Center for Axion and Precision Physics Research, Institute for Basic Science, Daejeon, Republic of Korea}
\author{Yannis K. Semertzidis}
\affiliation{Center for Axion and Precision Physics Research, Institute for Basic Science, Daejeon, Republic of Korea}
\affiliation{Department of Physics, Korea Advanced Institute of Science and Technology, Daejeon, Republic of Korea}
\author{Sergey Uchaikin}
\affiliation{Dark Matter Axion Group, Institute for Basic Science, Daejeon, Republic of Korea}
\affiliation{Center for Axion and Precision Physics Research, Institute for Basic Science, Daejeon, Republic of Korea}
\author{SungWoo Youn}
\email{swyoun@ibs.re.kr}
\affiliation{Dark Matter Axion Group, Institute for Basic Science, Daejeon, Republic of Korea}
\affiliation{Center for Axion and Precision Physics Research, Institute for Basic Science, Daejeon, Republic of Korea}

\date{\today}

\begin{abstract}
Conventional high-mass axion haloscopes based on the TM$_{010}$ mode lose detection volume as the resonant frequency increases.
We report an extended axion search using a dielectric-restored TM$_{020}$ cavity haloscope with symmetry-preserving auxetic tuning based on a single-degree-of-freedom mechanical architecture.
Using a near-quantum-limited microwave receiver, we searched a frequency range of 4.98--5.07\,GHz and exclude axion--photon couplings with sensitivity approaching the KSVZ benchmark.
Together with two earlier searches, the present scans extend a multi-scan program based on this architecture, yielding nearly 300\,MHz of contiguous high-mass axion coverage over 4.98--5.27 GHz, the first broad search reported with a single higher-order-mode haloscope.
This work establishes higher-order-mode cavities as a practical, scalable route beyond the TM$_{010}$ volume penalty.

\end{abstract}

\maketitle

The axion was proposed as a solution to the strong $CP$ problem~\cite{Peccei1977,Weinberg1978,Wilczek1978} and is a well-motivated dark matter candidate~\cite{Preskill1983,Abbott1983,Dine1983}.
A key experimental handle is its coupling to two photons, which enables axion-to-photon conversion in laboratory electromagnetic fields~\cite{Sikivie1983}.
The Kim--Shifman--Vainshtein--Zakharov (KSVZ)~\cite{Kim1979,Shifman1980} and Dine--Fischler--Srednicki--Zhitnitsky (DFSZ)~\cite{Dine1981,Zhitnitskii1980} models provide standard benchmark predictions for the axion--photon coupling and serve as key sensitivity targets for haloscope searches. 

In the microelectronvolt mass range, leading searches exploit this coupling with microwave cavity haloscopes, in which dark matter axions convert into photons in a resonant cavity immersed in a strong magnetic field.
For a cavity mode tuned to the axion Compton frequency, the expected conversion power scales as
\begin{equation}
P_{a\gamma\gamma} \propto g_{a\gamma\gamma}^{2}\rho_a \nu_a \langle B_e^2\rangle V_c Q_c C,
\label{eq:power_short}
\end{equation}
where $g_{a\gamma\gamma}$ is the axion--photon coupling, $\rho_a$ is the local dark matter density, $\nu_a$ is the axion Compton frequency, $\langle B_e^2\rangle$ is the volume-averaged square of the external magnetic field, $V_c$ and $Q_c$ are the cavity volume and unloaded quality factor, respectively.
The form factor is defined as
\begin{equation}
C = \frac{\left|\int \mathbf{E}_c\cdot\mathbf{B}_e\,dV\right|^2}
{\langle B_e^2\rangle V_c\int \epsilon_r |\mathbf{E}_c|^2 dV},
\label{eq:formfactor}
\end{equation}
where $\mathbf{E}_c$ is the cavity electric field, $\mathbf{B}_e$ is the external magnetic field, and $\epsilon_r$ is the relative permittivity.
It quantifies the overlap between the cavity mode and the external magnetic field, and sets the mode-dependent efficiency of axion-to-photon conversion.

Equations~\eqref{eq:power_short} and~\eqref{eq:formfactor} highlight the central challenge for high-frequency haloscopes: raising the resonant frequency without sacrificing $V_c$, $Q_c$, or $C$.
Conventional haloscopes favor the fundamental TM$_{010}$ mode of a cylindrical cavity, whose node-free electric field couples efficiently to a uniform solenoidal magnetic field.
However, the TM$_{010}$ resonant frequency scales inversely with cavity radius; pushing the search to higher axion masses requires smaller cavities, which rapidly reduces the detection volume and thus signal power.
This volume--frequency trade-off is a central obstacle for high-mass axion searches.

Several strategies have been proposed and developed to address this issue, including multi-cavity arrays~\cite{ADMX_multicav}, multi-cell resonators~\cite{Jeong2018}, dielectric structures~\cite{MADMAX2017}, wire-plasma haloscopes~\cite{Lawson2019}, and photonic-crystal resonators~\cite{Bae2023photonic}.
A complementary route is to use higher-order TM$_{0n0}$ modes, whose resonant frequencies increase with radial mode number without reducing the physical cavity volume~\cite{Kim2020, alesini2022search}.
These modes, however, face two long-standing challenges: 1) spatial phase reversals in the cavity electric field partially cancel the overlap with the external magnetic field, substantially reducing the axion form factor; and 2) practical frequency tuning requires preserving mode symmetry throughout the scan in order to avoid form-factor degradation and mode mixing.
These two challenges can be addressed simultaneously by combining dielectric field restoration with mechanically synchronized auxetic tuning. 

Our cavity architecture restores the form factor of higher-order modes by reshaping the electric-field distribution with strategically placed high-permittivity dielectric elements.
In the TM$_{020}$ implementation, a dielectric cylinder at the cavity center compresses the central field region and expands the outer field region, thereby enhancing the constructive overlap with the external magnetic field.
For a given cavity volume, the TM$_{020}$ mode resonates at 2.30 times the TM$_{010}$ frequency, providing access to higher axion masses without reducing the resonator volume. 
Electromagnetic simulations show that dielectric loading restores its form factor to as high as 0.55, more than four times the unloaded value and close to that of the TM$_{010}$ mode, as summarized in Table~\ref{tab:TM0n0_comp}. 
Together, these features directly address the conventional volume--frequency trade-off of TM$_{010}$ haloscopes. 

\begin{table}
\centering
\begin{tabular*}{\columnwidth}{@{\extracolsep{\fill}}c|ccccc@{}}
\toprule
~~~Mode~~~ & $\nu_c/\nu_c^{010}$ & $Q_c/Q_c^{010}$ & $V_c/V_c^{010}$ & $C$ & ~~~$C_{\rm diel}$~~~ \\
\midrule
TM$_{010}$ & 1 & 1 & 1 & 0.69 & -- \\
TM$_{020}$ & 2.30 & 1.51 & 1 & 0.13 & 0.55 \\
TM$_{030}$ & 3.60 & 1.90 & 1 & 0.05 & 0.45 \\
\bottomrule
\end{tabular*}
\caption{Simulated electromagnetic properties of representative TM$_{0n0}$ cavity modes for a cylindrical cavity.
The resonant frequency $\nu_c$, quality factor $Q_c$, and cavity volume $V_c$ are normalized to those of the TM$_{010}$ mode.
The form factors $C$ and $C_{\rm diel}$ correspond to the empty and dielectric-loaded cavities, respectively, with $C_{\rm diel}$ denoting the maximum value obtained with $\epsilon_r=9.7$.}
\label{tab:TM0n0_comp}
\end{table}

\begin{figure}
\centering
\includegraphics[width=0.75\linewidth]{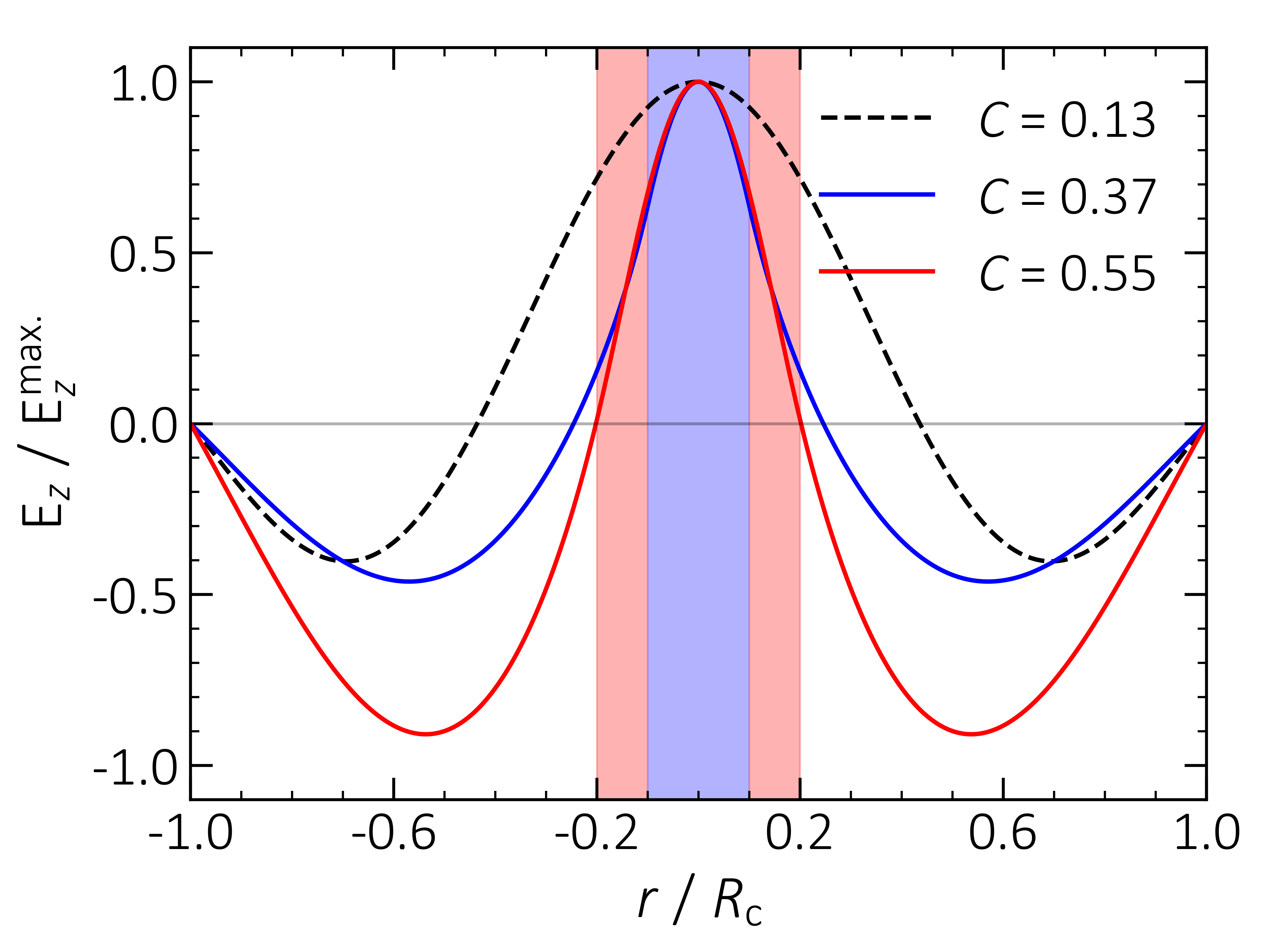}
\caption{Dielectric restoration of the TM$_{020}$ mode. 
The shaded regions indicate dielectric loading for two radial sizes, shown in blue and red. 
The corresponding normalized longitudinal electric-field profiles, $E_z/E_z^{\rm max}$, are plotted in the same colors as functions of the normalized radius $r/R_c$.
The black dashed curve shows the unloaded cavity for comparison.}
\label{fig:field_restore}
\end{figure}

The resonant frequency of dielectric-loaded TM$_{0n0}$ modes also highly sensitive to the dielectric size~\cite{Kim2020}, as reflected in the field distributions shown in Fig.~\ref{fig:field_restore}.
This frequency dependence motivates a tuning mechanism in which multiple dielectric rods are arranged symmetrically about the cavity center and displaced radially to vary the effective dielectric loading.
Because this motion is performed in a coordinated and symmetric manner, the cavity can be tuned continuously while preserving the azimuthal symmetry of the mode.

\begin{figure*}
\centering
\includegraphics[width=0.8\textwidth]{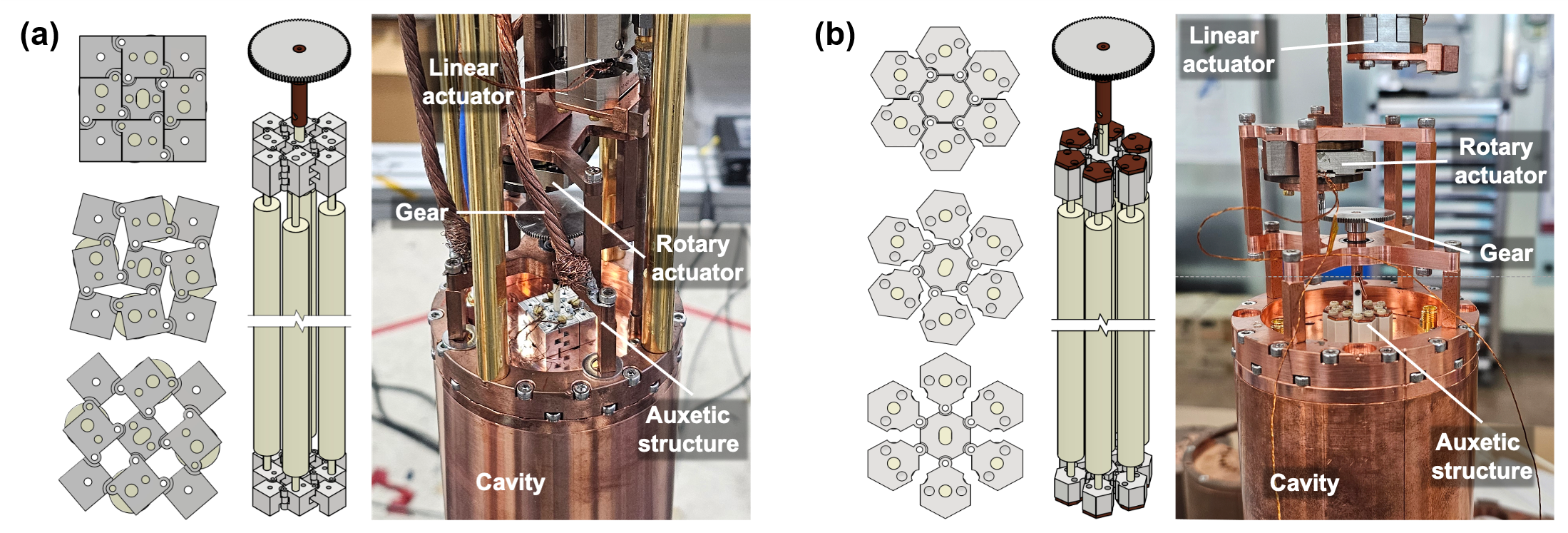}
\caption{Auxetic tuning structures for dielectric-loaded TM$_{020}$ cavities. 
(a) Square and (b) hexagonal configurations. 
For each configuration, the left panel illustrates the contracted and expanded states of the alumina-rod arrays, the middle panel shows a schematic view of the corresponding tuning assembly, and the right panel shows the fabricated system installed in the copper cavity. 
Rotation of the central shaft drives synchronized radial displacement of the dielectric rods, while the linear actuator adjusts the antenna coupling.\label{fig:auxetic}}
\end{figure*}

To meet this symmetry requirement, we use an auxetic mechanism in which the dielectric rods are kinematically linked through a kirigami-inspired structure.
Rotation of a central element drives coordinated radial displacement of the surrounding rods, producing uniform expansion or contraction of the array through a single control parameter, as illustrated in the left and middle panels of Fig.~\ref{fig:auxetic}.
This reduces a multi-degree-of-freedom tuning problem into single-degree-of-freedom actuation while preserving mode symmetry.
The auxetic linkage can be implemented in different geometries, allowing the tuning range and electromagnetic performance to be adjusted.
We employ square and hexagonal auxetic geometries that provide access complementary tuning ranges.
In the fabricated system, a cryogenic piezoelectric actuator coupled through a gear mechanism to the central dielectric element drives the synchronized radial motion, as shown in the right panels of Fig.~\ref{fig:auxetic}.

Building on our earlier searches with this architecture, referred to as Scans~1 and 2~\cite{Bae2024,Bae2025}, we report Scans~3 and 4, performed with the square and hexagonal auxetic configurations, respectively. 
These scans extend the search toward lower frequencies and bring the combined contiguous coverage to 4.98--5.27,GHz.
The new scans reach sensitivity approaching the KSVZ benchmark.

The experimental cavity was a cylindrical oxygen-free high-conductivity copper resonator with inner diameter 78\,mm and height 300\,mm.
It was designed to accommodate interchangeable endcaps carrying square and hexagonal dielectric-loaded auxetic tuning structures. 
In both configurations, high-purity alumina rods with relative permittivity $\epsilon_r=9.7$ formed the dielectric arrays.
The cavity dimensions and accessible dielectric spacings were chosen to provide continuous tuning across the frequency range needed to connect to and extend the previous TM$_{020}$ scans.
The square configuration allowed dielectric spacings from 6.5 to 9.2\,mm, while the hexagonal configuration covered 8.0 to 9.2\,mm.
Together, these configurations enabled continuous tuning over 4.88--5.29\,GHz.
The unloaded cavity quality factors remained above $10^5$ throughout the search range, with representative values of approximately $1.1\times10^5$ for Scan~3 and $1.2\times10^5$ for Scan~4.

\begin{figure}
\centering
\includegraphics[width=0.49\textwidth]{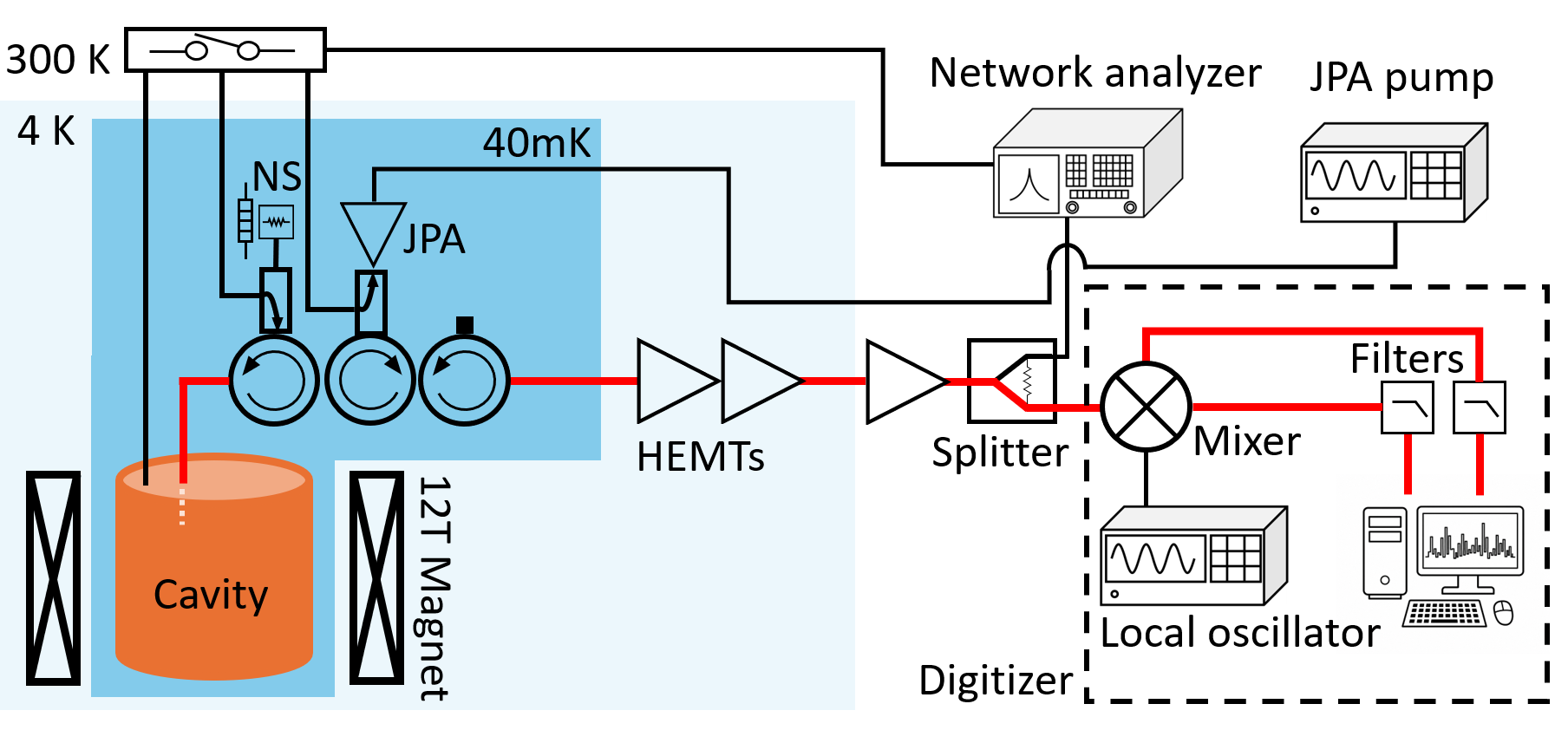}
\caption{Cryogenic haloscope setup. 
The dielectric-loaded TM$_{020}$ cavity is mounted at the 40-mK stage of a dilution refrigerator inside a 12-T superconducting solenoid magnet. 
The receiver chain consists of a JPA, circulators, cryogenic HEMTs, an RT down-conversion chain, and a digitizer, with the main DAQ path indicated in red. 
A calibrated noise source (NS) and a weakly coupled antenna are used for cavity and receiver characterization.\label{fig:setup}}
\end{figure}

The cavity was mounted at the 40-mK mixing-chamber stage of a dilution refrigerator and aligned with a 12-T superconducting solenoid magnet with a 96\,mm bore.
The volume-averaged magnetic field over the cavity was 9.8\,T.
Microwave signals were extracted from the cavity through a strongly coupled antenna and amplified by a near-quantum-limited, flux-tunable Josephson parametric amplifier (JPA) based on a dc-SQUID-terminated resonator, followed by cryogenic high-electron-mobility transistors (HEMTs) and room-temperature (RT) electronics.
The amplified signals were down-converted to an intermediate frequency of 3\,MHz and digitized at a sampling rate of 20\,MS/s.
Real-time Fourier transforms generated spectra over 1-MHz bandwidth with 100\,Hz resolution, much narrower than the expected virialized axion linewidth $\Delta\nu_a\sim \nu_a/Q_a\sim5$\,kHz at 5\,GHz for $Q_a\sim10^6$~\cite{Turner1990}.
The experimental setup is schematically shown in Fig.~\ref{fig:setup}.

The system noise temperature was measured {\it in situ} using the noise-visibility-ratio method~\cite{Sliwa2015}.
The output noise powers with the JPA on and off were compared as
$P_{\rm JPA,on}/P_{\rm JPA,off}=G_{\rm JPA}T_{\rm JPA,on}/T_{\rm JPA,off}$,
where $G_{\rm JPA}$ is the measured JPA gain, $T_{\rm JPA,on}$ corresponds to the system noise temperature, and $T_{\rm JPA,off}$ was independently determined from Y-factor calibrations.
This procedure was used to optimize the JPA operating point and monitor the receiver noise during the scans.

New data for Scan~3 and Scan~4 were acquired in 2025 and 2026 using the square and hexagonal auxetic tuning structures, respectively.
Scan~3 was conducted from 20 February to 6 May 2025, and Scan~4 from 2 September 2025 to 28 January 2026.
During these periods, periodic maintenance and occasional interruptions required system re-stabilization, including cavity characterization, antenna-coupling adjustment, and JPA re-optimization.

\begin{figure}
\centering
\includegraphics[width=0.75\linewidth]{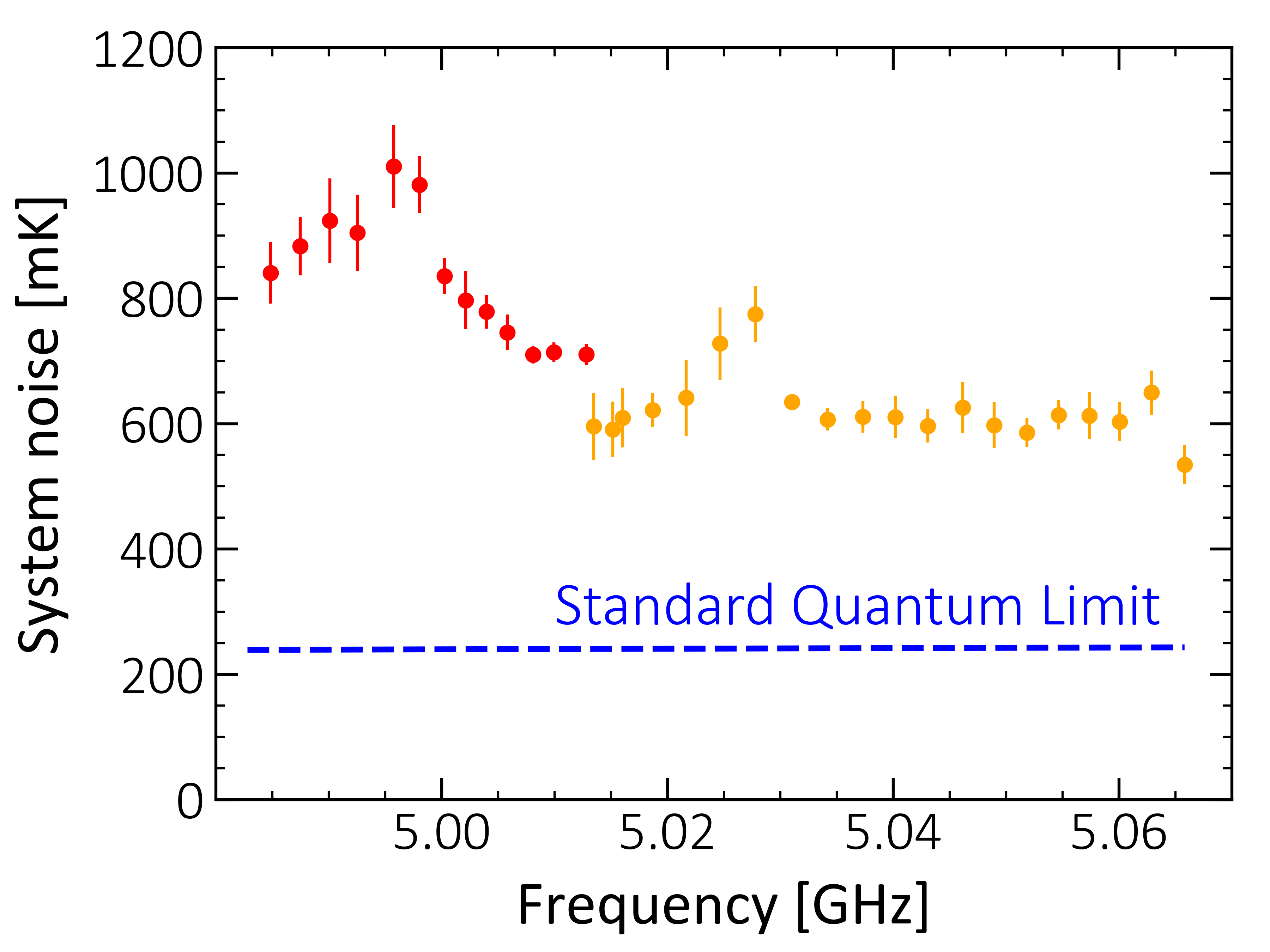}
\caption{Measured system noise temperature as a function of frequency for the square (orange) and hexagonal (red) configurations. 
The dashed blue line indicates the standard quantum limit.}
\label{fig:syst_noise}
\end{figure}

The data acquisition sequence was automated.
At each frequency step, the cavity resonance, quality factor, and antenna coupling were measured with a vector network analyzer.
The antenna coupling coefficient $\beta$ was extracted from a Smith-chart fit to the complex cavity reflection response and included in the calibration of the expected axion signal power.
The JPA operating point was optimized by minimizing the system noise temperature using the Nelder--Mead algorithm~\cite{NelderMead1965}, with the flux-bias current and pump power treated as free parameters.
The resulting system noise temperature was typically 600\,mK for Scan~3 and 850\,mK for Scan~4, as shown in Fig.~\ref{fig:syst_noise}, corresponding to about 2.5 and 3.5 noise photons at 5\,GHz.
Power spectra were then acquired over the prescribed integration time: 30\,min per frequency step for Scan~3 and 90\,min for Scan~4, with the longer integration compensating for the lower form factor and higher noise temperature.
After each acquisition, the cavity was tuned by 30\,kHz toward lower frequencies, and the automated sequence was repeated.

The obtained power spectra were analyzed with the standard haloscope procedure described in Ref.~\cite{Brubaker2017}.
Frequency-dependent baselines from gain variations and impedance mismatches were removed with a Savitzky--Golay filter~\cite{Savitzky1964}.
The baseline-subtracted spectra were rescaled by the expected Lorentzian cavity response and measured system noise to place the individual power excesses on a common SNR-equivalent scale.
Spectral bins corresponding to the same axion frequency were then vertically combined across tuning steps using inverse-variance weighting, yielding a single combined spectrum over the full search range.
The combined spectrum was subsequently processed with a matched filter based on the Standard Halo Model (SHM) axion lineshape, which accounts for signal broadening arising from the Galactic velocity distribution of dark matter~\cite{Turner1990}.
This produced the grand spectrum with each bin representing possible axion-induced excess power at the corresponding trial masses.

The analysis efficiency was evaluated using synthetically generated axion signals.
Monte Carlo signals with a reference SNR of 5 were injected into the raw spectra, which were then processed through the full analysis chain.
The recovered SNR was compared with the injected value to quantify the signal attenuation introduced by the analysis procedure.
This yielded average SNR efficiencies of $\epsilon_{\rm SNR}=86.8\%$ for Scan~3 and $82.2\%$ for Scan~4.

\begin{figure*}
\includegraphics[width=0.75\linewidth]{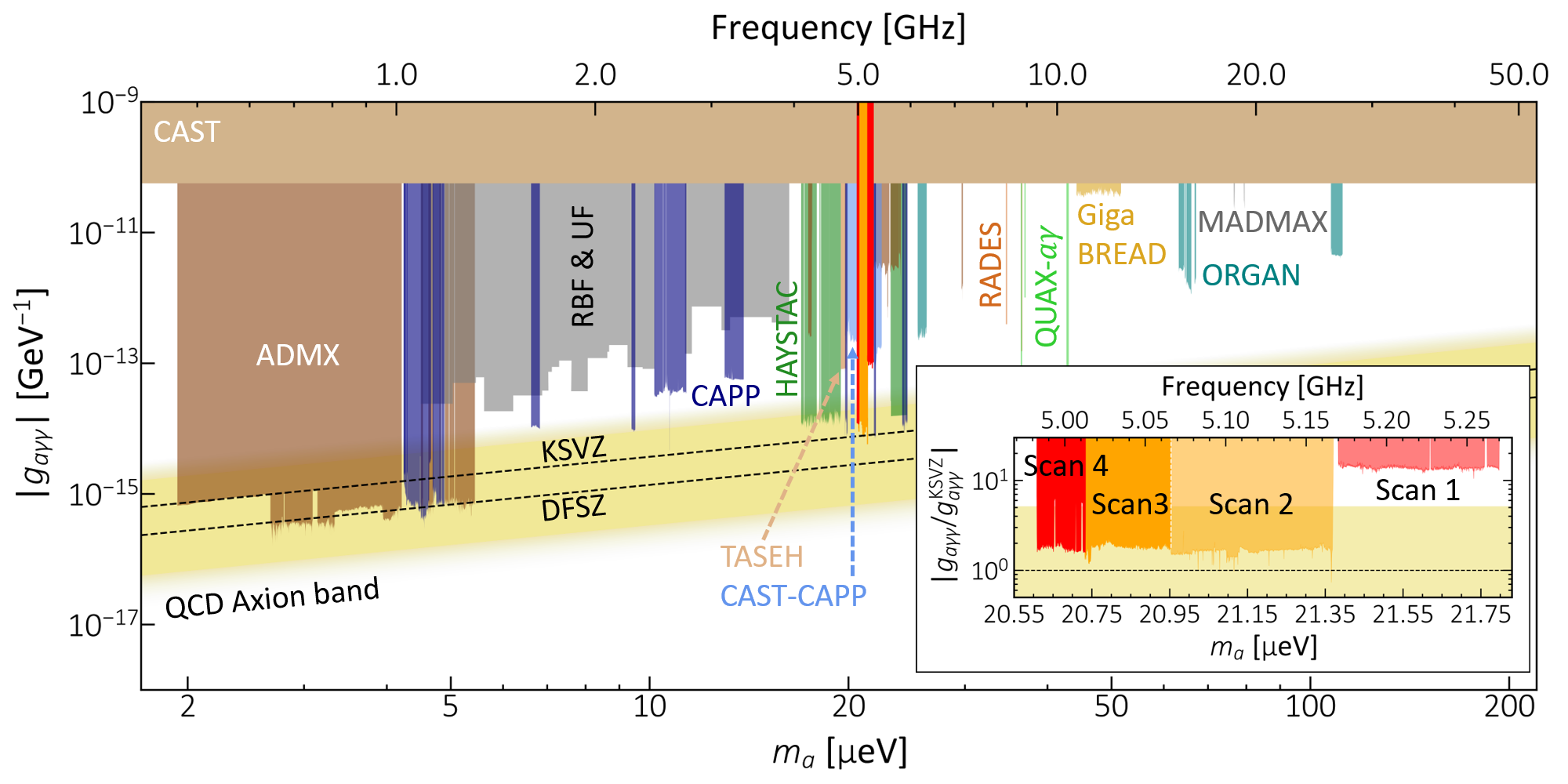}
\caption{Exclusion limits on the axion--photon coupling from the dielectric-loaded TM$_{020}$ searches.
The yellow band denotes the representative QCD axion models.
The present scans, Scans~3 and 4, cover 4.98--5.07\,GHz with sensitivity approaching the KSVZ benchmark.
Combined with previous Scans~1 and 2~\cite{Bae2024,Bae2025}, they provide contiguous coverage from 4.98 to 5.27\,GHz, corresponding to axion masses of 20.61--21.79\,$\mu$eV.
The inset shows a magnified view of this region, with the hexagonal and square configurations shown in red and orange, respectively.
Existing cavity-haloscope limits are shown for comparison~\cite{OHare2020}.}
\label{fig:limits}
\end{figure*}

The grand spectra were normalized by the measured noise standard deviation before candidate selection.
Bins with normalized excess power above 3.719 were selected as candidates, a threshold chosen so that an SNR-5 axion signal would cross it with 90\% probability.
All candidate frequencies were rescanned with additional integration time and reanalyzed with the same pipeline.
No candidate excess remained above threshold after the rescan procedure.
We therefore set 90\% confidence-level upper limits on $g_{a\gamma\gamma}$ assuming the SHM lineshape and a local dark-matter density of $\rho_a=0.45\,\mathrm{GeV/cm^3}$~\cite{deSalas2021,Ou2024,Lim2025}.
The two new scans cover 4.98--5.07\,GHz, corresponding to axion masses of 20.61--20.95\,$\mu$eV, and reach an average sensitivity of approximately $1.8\times g_{a\gamma\gamma}^{\rm KSVZ}$.
Figure~\ref{fig:limits} shows the resulting limits together with previous cavity-haloscope constraints.

The exclusion limits included uncertainties in the analysis efficiency, system noise temperature, cavity form factor, antenna coupling, and cavity quality factor, as summarized in Table~\ref{tab:uncertainties}.
The dominant contributions arose from $\epsilon_{\rm SNR}$, reflecting the frequency-dependent variation of injected-signal recovery, and $T_{\rm sys}$, which includes uncertainties in the measured JPA gain and the independent calibration of $T_{\rm JPA,off}$.
The form-factor uncertainty was estimated from simulations of tuning-rod misalignment, while those in $\beta$ and $Q_c$ were obtained from the Smith-chart fit and statistical fluctuations in the cavity characterization, respectively.
All contributions were propagated into the final axion--photon coupling limits.

\begin{table}
\centering
\begin{tabular*}{\columnwidth}{@{\extracolsep{\fill}}cccc@{}}
\toprule
\multirow{2}{*}{Parameter}
& \multicolumn{2}{c}{Uncertainty [\%]}
& \multirow{2}{*}{Source} \\
\cmidrule(lr){2-3}
& Scan~3 & Scan~4 & \\
\midrule
$\epsilon_{\rm SNR}$ & 6.5 & 4.4 & Signal-recovery variation \\
$T_{\rm sys}$ & 6.1 & 5.2 & $G_{\rm JPA}$ and $T_{\rm JPA,off}$ calibration \\
$C$ & 3.5 & 4.5 & Tuning-rod misalignment \\
$\beta$ & 1.1 & 1.9 & Smith-chart fit \\
$Q_c$ & $<1$ & 2.0 & Statistical fluctuation \\
\bottomrule
\end{tabular*}
\caption{Uncertainty budget for Scan~3 and Scan~4.
All values are given as percentages.}
\label{tab:uncertainties}
\end{table}

The series of four auxetically tuned TM$_{020}$ searches provides contiguous coverage from 4.98 to 5.27\,GHz, corresponding to nearly 300\,MHz of high-mass axion coverage within a single higher-order-mode architecture. 
This capability arises from combining dielectric field restoration with symmetry-preserving auxetic tuning, enabling large effective volume and continuous frequency tuning through single-degree-of-freedom actuation. 
These results establish dielectric-restored higher-order-mode cavities as practical resonators for broadband high-mass axion searches. 
The same principle can be extended to higher TM$_{0n0}$ modes, providing access to still higher axion masses without the severe volume penalty of conventional TM$_{010}$ cavities.

\begin{acknowledgments}
This work was supported by the Institute for Basic Science (Grant Nos. IBS-R040-C1 and IBS-R017-D1) and by JSPS KAKENHI (Grant No. JP22H04937).
J.J. was supported by the Knut and Alice Wallenberg Foundation.
Y.K. was supported by the Alexander von Humboldt Foundation.
A.F.v.L. was supported by a JSPS Postdoctoral Fellowship.
J.E.K. was supported by the National Academy of Sciences of the Republic of Korea.
\end{acknowledgments}

\bibliographystyle{apsrev4-2}
\bibliography{mybib}

@article{Peccei1977,
  author  = {Peccei, R. D. and Quinn, Helen R.},
  title   = {{{CP} conservation in the presence of pseudoparticles}},
  journal = {Phys. Rev. Lett.},
  volume  = {38},
  pages   = {1440--1443},
  year    = {1977}
}

@article{Weinberg1978,
  author  = {Weinberg, Steven},
  title   = {A new light boson?},
  journal = {Phys. Rev. Lett.},
  volume  = {40},
  pages   = {223--226},
  year    = {1978}
}

@article{Wilczek1978,
  author  = {Wilczek, Frank},
  title   = {Problem of strong {$P$} and {$T$} invariance in the presence of instantons},
  journal = {Phys. Rev. Lett.},
  volume  = {40},
  pages   = {279--282},
  year    = {1978}
}

@article{Preskill1983,
  author  = {John Preskill and Mark B. Wise and Frank Wilczek},
  title   = {Cosmology of the invisible axion},
  journal = {Phys. Lett. B},
  volume  = {120},
  pages   = {127--132},
  year    = {1983}
}

@article{Abbott1983,
  author  = {L. F. Abbott and P. Sikivie},
  title   = {A cosmological bound on the invisible axion},
  journal = {Phys. Lett. B},
  volume  = {120},
  pages   = {133--136},
  year    = {1983}
}

@article{Dine1983,
  author  = {Michael Dine and Willy Fischler},
  title   = {The not-so-harmless axion},
  journal = {Phys. Lett. B},
  volume  = {120},
  pages   = {137--141},
  year    = {1983}
}

@article{Kim1979,
  author  = {Kim, Jihn E},
  title   = {Weak-interaction singlet and strong {CP} invariance},
  journal = {Phys. Rev. Lett.},
  volume  = {43},
  pages   = {103},
  year    = {1979}
}

@article{Shifman1980,
  author  = {Shifman, Mikhail A and Vainshtein, AI and Zakharov, Valentin I},
  title   = {Can confinement ensure natural {CP} invariance of strong interactions?},
  journal = {Nucl. Phys. B},
  volume  = {166},
  pages   = {493--506},
  year    = {1980}
}

@article{Dine1981,
  author  = {Dine, Michael and Fischler, Willy and Srednicki, Mark},
  title   = {A simple solution to the strong {CP} problem with a harmless axion},
  journal = {Phys. Lett. B},
  volume  = {104},
  pages   = {199--202},
  year    = {1981}
}

@article{Zhitnitskii1980,
  author  = {Zhitnitskii, AP},
  title   = {Possible suppression of axion--hadron interactions},
  journal = {Sov. J. Nucl. Phys.(Engl. Transl.);(United States)},
  volume  = {31},
  year    = {1980}
}

@article{Sikivie1983,
  author  = {Sikivie, P.},
  title   = {Experimental tests of the ``invisible'' axion},
  journal = {Phys. Rev. Lett.},
  volume  = {51},
  pages   = {1415--1417},
  year    = {1983}
}

@phdthesis{ADMX_multicav,
  author = {Kinion, D. S.},
  title  = {First results from a multiple microwave cavity search for dark matter axions},
  school = {University of California, Davis},
  year   = {2001}
}

@article{MADMAX2017,
  title = {Dielectric Haloscopes: A New Way to Detect Axion Dark Matter},
  author = {Caldwell, Allen and Dvali, Gia and Majorovits, B\'ela and Millar, Alexander and Raffelt, Georg and Redondo, Javier and Reimann, Olaf and Simon, Frank and Steffen, Frank},
  collaboration = {MADMAX Working Group},
  journal = {Phys. Rev. Lett.},
  volume = {118},
  issue = {9},
  pages = {091801},
  numpages = {6},
  year = {2017},
  month = {Mar},
  publisher = {American Physical Society}
}

@article{Jeong2018,
  author  = {Junu Jeong and SungWoo Youn and Saebyeok Ahn and Jihn E. Kim and Yannis K. Semertzidis},
  title   = {Concept of multiple-cell cavity for axion dark matter search},
  journal = {Phys. Lett. B},
  volume  = {777},
  pages   = {412--419},
  year    = {2018}
}

@article{Lawson2019,
  author  = {Lawson, Matthew and Millar, Alexander J. and Pancaldi, Matteo and Vitagliano, Edoardo and Wilczek, Frank},
  title   = {Tunable axion plasma haloscopes},
  journal = {Phys. Rev. Lett.},
  volume  = {123},
  pages   = {141802},
  year    = {2019}
}

@article{Bae2023photonic,
  author  = {Bae, Sungjae and Youn, SungWoo and Jeong, Junu},
  title   = {Tunable photonic crystal haloscope for high-mass axion searches},
  journal = {Phys. Rev. D},
  volume  = {107},
  pages   = {015012},
  year    = {2023}
}

@article{Kim2020,
  author  = {Jinsu Kim and SungWoo Youn and Junu Jeong and Woohyun Chung and Ohjoon Kwon and Yannis K Semertzidis},
  title   = {Exploiting higher-order resonant modes for axion haloscopes},
  journal = {J. Phys. G Nucl. Part. Phys.},
  volume  = {47},
  pages   = {035203},
  year    = {2020}
}

@article{Bae2024,
  author  = {Bae, Sungjae and Jeong, Junu and Kim, Younggeun and Youn, SungWoo and Park, Heejun and Seong, Taehyeon and Oh, Seongjeong and Semertzidis, Yannis K.},
  title   = {Search for dark matter axions with tunable {${\mathrm{TM}}_{020}$} mode},
  journal = {Phys. Rev. Lett.},
  volume  = {133},
  pages   = {211803},
  year    = {2024}
}

@article{Bae2025,
  author  = {Bae, Sungjae and Jeong, Junu and Kim, Younggeun and Youn, SungWoo and Kim, Jinsu and van Loo, Arjan F. and Nakamura, Yasunobu and Oh, Seonjeong and Seong, Taehyeon and Uchaikin, Sergey and Kim, Jihn E. and Semertzidis, Yannis K.},
  title   = {Axion dark matter search with sensitivity near the {Kim--Shifman--Vainshtein--Zakharov} benchmark using the {${\mathrm{TM}}_{020}$} mode},
  journal = {Phys. Rev. Lett.},
  volume  = {135},
  pages   = {091804},
  year    = {2025}
}

@article{alesini2022search,
  author  = {Alesini, D. and Babusci, D. and Braggio, C. and Carugno, G. and Crescini, N. and D'Agostino, D. and D'Elia, A. and Di Gioacchino, D. and Di Vora, R. and Falferi, P. and others},
  title   = {Search for Galactic axions with a high-$Q$ dielectric cavity},
  journal = {Phys. Rev. D},
  volume  = {106},
  pages   = {052007},
  year    = {2022}
}

@article{Sliwa2015,
  author  = {Sliwa, K. M. and Hatridge, M. and Narla, A. and Shankar, S. and Frunzio, L. and Schoelkopf, R. J. and Devoret, M. H.},
  title   = {Reconfigurable Josephson circulator/directional amplifier},
  journal = {Phys. Rev. X},
  volume  = {5},
  pages   = {041020},
  year    = {2015}
}

@article{Ou2024,
  author  = {Ou, Xiaowei and Eilers, Anna-Christina and Necib, Lina and Frebel, Anna},
  title   = {The dark matter profile of the Milky Way inferred from its circular velocity curve},
  journal = {Mon. Not. R. Astron. Soc.},
  volume  = {528},
  pages   = {693--710},
  year    = {2024}
}

@article{deSalas2021,
  author  = {de Salas, P. F. and Widmark, A.},
  title   = {Dark matter local density determination: Recent observations and future prospects},
  journal = {Rep. Prog. Phys.},
  volume  = {84},
  pages   = {104901},
  year    = {2021}
}

@article{Lim2025,
  author  = {Lim, Sung Hak and Putney, Eric and Buckley, Matthew R. and Shih, David},
  title   = {Mapping dark matter in the Milky Way using normalizing flows and Gaia DR3},
  journal = {J. Cosmol. Astropart. Phys.},
  volume  = {2025},
  pages   = {021},
  year    = {2025}
}

@article{NelderMead1965,
  author  = {Nelder, John A and Mead, Roger},
  title   = {A simplex method for function minimization},
  journal = {Comput. J.},
  volume  = {7},
  pages   = {308--313},
  year    = {1965}
}

@article{Savitzky1964,
  author  = {Savitzky, Abraham and Golay, Marcel JE},
  title   = {Smoothing and differentiation of data by simplified least squares procedures.},
  journal = {Anal. Chem.},
  volume  = {36},
  pages   = {1627--1639},
  year    = {1964}
}

@article{Turner1990,
  author  = {Turner, Michael S},
  title   = {Periodic signatures for the detection of cosmic axions},
  journal = {Phys. Rev. D},
  volume  = {42},
  pages   = {3572},
  year    = {1990}
}

@article{Brubaker2017,
  author  = {Brubaker, BM and Zhong, L and Lamoreaux, SK and Lehnert, KW and Van Bibber, KA},
  title   = {HAYSTAC axion search analysis procedure},
  journal = {Phys. Rev. D},
  volume  = {96},
  pages   = {123008},
  year    = {2017}
}

@misc{OHare2020,
  author    = {O'Hare, Ciaran A. J.},
  title     = {{AxionLimits}},
  publisher = {Zenodo},
  version   = {v1.0},
  year      = {2020}
}

\end{document}